\documentstyle[aps,multicol,epsfig]{revtex}
\begin{document}
\draft
\title{Attractive forces between circular polyions \\
of the same charge}
\author{Silvia Martins and J\"urgen F. Stilck}
\address{Instituto de F\'{\i}sica, Universidade Federal Fluminense\\
Av. Litor\^anea s/n, 24210-340, Niter\'oi, RJ, Brazil}
\date{\today}
\maketitle
\begin{abstract}
We study two models of ringlike polyions which are two-dimensional versions of
simple models for colloidal particles (model A) and for rodlike segments of
DNA (model B), both in solution with counterions. The counterions may
condense on $Z$ sites of the polyions, and we suppose the number of
condensed counterions on each polyion $n$ to be fixed. The exact free energy of a
pair of polyions is calculated for not too large values of $Z$, and for both
models we find that attractive forces appear between the rings even when the
condensed counterions do not neutralize the total charge of the polyions. This
force is due to correlations between the condensed counterions and in general
becomes smaller as the temperature is increased. For model A divergent force
may appear as the separation between the rings vanishes, and this makes an
analytical study possible for this model and for vanishing separation, showing a
universal behavior in this limit. Attractive forces are found for model A if
the valence of the counterions is larger than one. For model B, no such
divergences are present, and attractive forces are found for a finite range of
values of the counterion valence, which depends of $Z$, $n$, and the
temperature. 
\end{abstract}


\begin{multicols}{2}
\section{Introduction}
\label{I}

The interaction between colloidal particles in suspensions are a
statistical mechanical problem which has attracted much attention
from both theoretical and experimental points of view for more than
one century now, but there are still issues which are not totally
understood \cite{gc2000}. Besides the difficulty of properly taking
into account the long range Coulomb interactions, it is necessary to consider
the fact that each
colloidal particle in solution may contribute with a quite large
number of counterions, and their contribution to the thermodynamic
properties of the suspension is quite important \cite{lbt98}. One
particularly important issue related to these systems is the
possibility of phase separation, and although it has
been argued that phase separation may occur even in situations where
{\em repulsive} interactions between pairs of colloidal particles are
present \cite{rh97}, recently it was found that for aqueous solutions
with monovalent counterions no phase separation is expected
\cite{dbl01} and this seems to be consistent with conclusions drawn
from simulations \cite{ll99}. For {\em multivalent} counterions,
however, correlations between counterions which are condensed on the
colloidal particles may produce short range {\em attractive}
effective interactions between the particles, even if their bare
charge is not totally neutralized \cite{sr90}. This interesting
phenomenon of effective attractive interactions between objects with
charges of the same sign has attracted much attention recently,
particularly for systems of rodlike polyions, such as segments of DNA
\cite{blo91}

The theoretical approaches used so far to address the problem of
attraction between rodlike polyions of the same charge may be
classified in two groups. The first formulates the system as a
field-theoretic problem and then studies its properties in a high
temperature approximation \cite{ha97}. The second approach, which is also
approximate, supposes the system of polyions and condensed
counterions to form a Wigner crystal at zero temperature, and then
proceeds considering the effect of temperature on the system
\cite{rou96,are99,sol99,kor99,lev99}.
Some time ago, a very simple model has been proposed for an isolated
pair of rodlike polyions in a solution with counterions which may be
exactly solved, with the aid of a computer, for not too large rods
\cite{are99}. In this model, the only effect of the condensation of a
counterion on the site where an ion is located is to renormalize the
local charge. The valence of each counterion is $\alpha$, so that each
counterion has a
charge $\alpha q$, and the polyion consist of a linear array of $Z$ charges
$-q$ with a
distance $b$ between first neighbors. The solution has an effective
dielectric constant
$D$. The mean value of the number $n$ of condensed counterions may
be determined by the Manning criterion \cite{man69} as
\begin{equation}
n=\left(1-\frac{1}{\alpha\xi}\right)\frac{Z}{\alpha}
\label{man}
\end{equation}
where
\begin{equation}
\xi=\frac{q^2}{Dbk_BT}
\end{equation}
is the Manning parameter. For sufficiently low temperature ($\xi>1/\alpha$),
counterions will therefore condense on the polyions. It should be remarked
that the expression
above for the number of condensed counterions is an approximation for
an isolated polyion and in the limit $Z \to \infty$. One may wonder
if it is still valid for finite values of $Z$ and, supposing that the
Manning value for $n$ is the mean value of a {\em distribution} of
numbers of condensed counterions it is relevant to ask if this
distribution is broad or narrow. It turns out that the expression
(\ref{man}) is actually a good approximation even for rather small values
of $Z$ and that in usual physical conditions the distribution of
values of $n$ is quite narrow \cite{lb97}. Now if we consider two
polyions separated by a distance $d$, we may use expression (\ref{man}) in
the limits $d \to \infty$ and $d \to 0$. The total number of
counterions condensed on both rods in the limit $d \to \infty$ will be
\begin{equation}
n_{\infty}=2n
\end{equation}
and if we now consider the other limit $d \to 0$ (where both chains
merge into one with $Z$ charges $-2q$) the number of condensed
charges will now be given by
\begin{equation}
n_0=\left(1-\frac{1}{2\alpha\xi}\right)\frac{2Z}{\alpha}
\end{equation}
and therefore the change in the number of condensed counterions is
given by
\begin{equation}
\Delta n=n_0-n_{\infty}=n_{\infty}\frac{1}{\alpha\xi-1}
\end{equation}
and for not too high temperatures it may be a reasonable
approximation to ignore this change, as is done in the simple model
we described.

In this paper we study two models for the interaction between two
ringlike charge structures which are inspired by studies of colloidal
particles and of rodlike polyions, respectively. The models are similar to the 
one  for
rodlike polyions described above, and we study the thermodynamic
properties of these models by solving them exactly for not too large
particles. In the first model (model A), the bare colloidal particles are 
supposed
to be formed by $Z$ point charges $-q$ placed regularly on circles, and we
consider two of these ringlike particles placed in a solution with
counterions. The number of counterions condensed on each particle is
equal to $n$ and we proceed calculating the force between them as a
function of the separation between the rings, for fixed values of the
valence $\alpha$ of the counterions and the Manning parameter $\xi$.
The second model is a two-dimensional version for a model which was recently
proposed 
for rodlike DNA segments, in which the negative charge of the polyion is 
supposed to be concentrated at the axis of a cylinder, while the counterions 
are condensed at the surface of this cylinder \cite{dl01}. Thus, in model B we
each ringlike polyion is formed by a point charge $-Zq$ in the center 
of the circle, while $n$ counterions may condense on $Z$ equally spaced sites 
on the circle. The basic point we address in our calculation is the possibility 
of attractive forces between polyions of the same total charge, and for both 
models we find that such forces are indeed possible.

In the next section we present both models in detail and the results we
obtained for them, which are either analytical or numerically exact. Final
comments and discussions may be found in section \ref{III}.

\section{Definition of the models and their solution}
\label{II}

\begin{figure}
\centerline{\epsfig{file=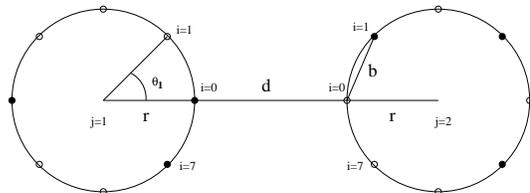,width=7cm} }
\caption{A possible configuration of two ring polyions with $Z=8$ and $n=3$
condensed counterions. Full small circles indicate adsorption sites with 
condensed counterions, whereas empty small circles are empty condensation 
sites.}
\label{f1}
\end{figure}

We consider two circular polyions of radius $r$ with an even number $Z$ of
point charges $-q$ which are placed as follows:
\begin{itemize}
\item{Model A} Uniformly distributed on the circle.
\item{Model B} In the center of the circle.
\end{itemize}
These polyions are in a ionic solution of
$\alpha$-valent counterions, also supposed to be point charges (charge $\alpha
q$). We will admit that $n$ of the counterions are condensed on $Z$ sites on 
the circle. In model A, since the counterions condense on top of the fixed 
negative charges, the only effect of a condensed counterion is to renormalize 
the charge on
the condensation site, which will be equal to $q(\alpha-1)$. In model B, the 
charge at an condensation site on the circle is equal to zero if no counterion 
is condensed on it and $\alpha q$ otherwise. The distance
between the centers of the polyions is $2r+d$. The definitions above are
illustrated in figure \ref{f1}.

The positions of the condensation sites on the rings will be referred by two 
indices
$j=1,2$, indicating the ring, and $i=0,1,\ldots,Z-1$, indicating the position
on the ring. For model A, the electrostatic interactions between he charges may 
then be
described by the Hamiltonian
\begin{equation}
{\cal H}_A=\frac{q^2}{2Dr}\sum_{i,i^{'}=0}^{Z-1}\sum_{j,j^{'}=1}^2
\frac{(1-\alpha \sigma_{i,j})(1-\alpha \sigma_{i^{'},j^{'}})} 
{\delta(i,j,i^{'},j^{'})},
\label{e1}
\end{equation}
where the sum is restricted to $(i,j) \ne (i^\prime,j^\prime)$ and the
occupation 
variables $\sigma_{i,j}$ assume the value 0 if no counterion is condensed on
site $(i,j)$ and 1 otherwise. $D$ is the dielectric constant and the ratio of
the distance between two sites and $r$ is given by
\begin{eqnarray}
\lefteqn{\delta(i,j;i^{\prime},j^{\prime})= } \\
& &\left\{
\begin{array}{ll}
2 \left| \sin \frac{\theta_{i,i^\prime}}{2}\right| & \mbox{if
$j=j^{\prime}$,}\nonumber\\ 
\sqrt{(x+2-cos\theta_i-cos\theta_{i^{\prime}})^2+
(\sin\theta_i-\sin\theta_{i^{\prime}})^2} & \mbox{otherwise;}
\end{array}
\right.
\label{dist}
\end{eqnarray}
where $x=d/r$ and 
\begin{eqnarray}
\theta_{i,i^\prime}=\frac{2\pi(i-i^\prime)}{Z},\\
\theta_i=\theta_{i,0}.
\end{eqnarray}

Model B may be described by the Hamiltonian
\begin{eqnarray}
{\cal H}_B=\frac{q^2\alpha^2}{2Dr}\sum_{i,i^{'}=0}^{Z-1}\sum_{j,j^{'}=1}^2
\frac{\sigma_{i,j}\sigma_{i^{'},j^{'}}} 
{\delta(i,j,i^{'},j^{'})}- \nonumber \\
\frac{Zq^2\alpha}{Dr}\sum_{i=0}^{Z-1}\frac{\sigma_{i,1}+\sigma_{i,2}} 
{\delta(i)}+\frac{Z^2q^2}{r}\frac{1}{2+x},
\label{e1b}
\end{eqnarray}
where the interaction between the central charge and the condensed counterions 
on the same ring was not included, since it is constant, and
\begin{equation}
\delta(i)=\sqrt{1+(2+x)^2-2(2+x)\cos\theta_i}
\label{distb}
\end{equation}

The partition function $Q$ is calculated summing over the possible
positions of the condensed counterions
\begin{equation}
Q_{A,B}={\sum_{\{\sigma_{i,j}\}}}^{\prime}\exp(-\beta{\cal H}_{A,B}),
\label{e3}
\end{equation}
where the prime denotes the constraint of having $n$ counterions condensed on
each ring.

It is convenient to define adimensional hamiltonians $H$ through
\begin{equation}
\beta{\cal H}_{A,B}=\xi^\prime H_{A,B},
\end{equation}
where $\xi^\prime=\beta q^2/Dr$ is related to the Manning
parameter defined in expression (\ref{man}) for rodlike polyions. Then 
\begin{equation}
H_A=\frac{1}{2}\sum_{(i,j) \neq (i^{'},j^{'})}
\frac{(1-\alpha \sigma_{ij})(1-\alpha \sigma_{i^{'}j^{'}})} 
{\delta(i,j,i^{'},j^{'})},
\label{e5}
\end{equation}
and
\begin{eqnarray}
H_B=\frac{\alpha^2}{2}\sum_{(i,j) \neq (i^{'},j^{'})}
\frac{\sigma_{i,j}\sigma_{i^{'},j^{'}}} 
{\delta(i,j,i^{'},j^{'})}\nonumber \\
-Z\alpha\sum_i\frac{\sigma_{i,1}+\sigma_{i,2}}{\delta(i)}+\frac{Z^2}{2+x}.
\label{e5b}
\end{eqnarray}

We may define the statistical weights
\begin{equation}
y_i=\exp\left(\frac{-\xi^\prime}{a_i}\right),i=1,2,\ldots,Z/2,
\end{equation}
and
\begin{equation}
z_i=\exp\left(\frac{-\xi^\prime}{b_i}\right),i=1,2,\ldots,m,
\end{equation}
related to the intra- and interpolyion interactions between charges at the 
condensation sites, respectively, and
\begin{equation}
t_i=\exp\left(\frac{-\xi^\prime}{c_i}\right),i=0,1,\ldots,Z/2,
\end{equation}
which correspond to the interaction between a central charge of a ring and the 
condensed counterions on the other ring.
The denominators in the arguments of the exponentials are the adimensional
distances between the interacting charges. For a pair of charges in the same
polyion
\begin{equation}
a_i=2\left| \sin \frac{\theta_i}{2} \right|.
\end{equation}
The expressions
for the denominators $b_i$ are straightforward to obtain from (\ref{dist}) but 
will be omitted
here for brevity. $m+1=Z^2/4+Z/2+1$ is the number
of different distances between condensation sites located on distinct rings. 
Finally, the adimensional distances between condensation sites and the central 
charge of the other ring are $c_i=\delta_i$, according to equation (\ref{distb}). 

Using these statistical weights, we may rewrite the partition function as 
\begin{equation}
Q_{A,B}=\sum_{i=1}^{N_c}\omega_{A,B}(i)
\end{equation}
where the sum runs over the
\begin{equation}
N_c=\left[\frac{Z!}{n!(Z-n)!)}\right]^2
\end{equation}
allowed configurations of condensed counterions and
\begin{equation}
\omega_A(i)=\prod_{j=1}^{Z/2}y_j^{u_A(i,j)}\prod_{k=0}^m z_k^{v_A(i,k)}.
\end{equation}
The exponents $u_A(ij)$ and $v_A(ik)$ are quadratic polynomials in $\alpha$ 
with 
integer coefficients. For model B we obtain
\begin{equation}
\omega_B(i)=\prod_{j=1}^{Z/2}y_j^{u_B(i,j)}\prod_{k=0}^m z_k^{v_B(i,k)} 
\prod_{l=0}^{Z/2} t_l^{w_B(i,l)},
\end{equation}
where $u_B(ij)$ and $v_B(ik)$ are integer coefficients multiplied by $\alpha^2$ 
and $w_B(i,l)$ are negative integer coefficients multiplied by $Z\alpha$. 
The whole set of integer coefficients can be calculated
with the help of a
computer program, and then we can obtain the partition function exactly for 
both models and not too large values of $Z$.

The force between the rings is given by
\begin{equation}
F_{A,B}=-\frac{\partial \phi_{A,B}}{\partial d},
\end{equation}
where $\phi_{A,B}=-k_B T\ln Q_{A,B}$ is the Helmholtz free-energy. It is 
convenient to define an adimensional force $f_{A,B}=(Dr^2/q^2)F_{A,B}$, and 
thus 
\begin{equation}
f_{A,B}=\frac{1}{\xi^\prime Q_{A,B}}\frac{\partial Q_{A,B}}{\partial x}.
\end{equation}

\begin{figure}
\centerline{\epsfig{file=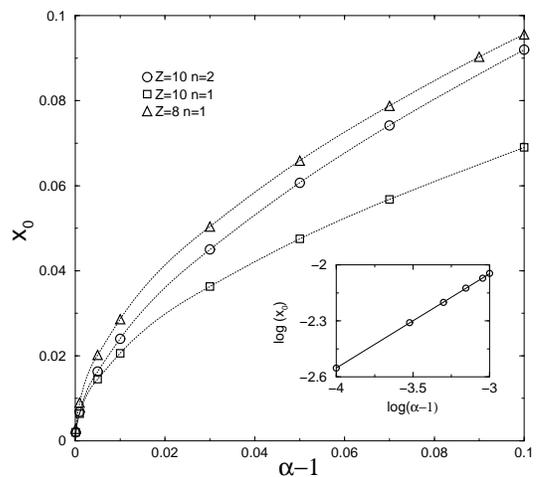,width=7cm} }
\caption{The equilibrium distance $x_0$ as a function of $\alpha-1$ for
$\xi^\prime=2$ (model A). The inset shows data for $Z=8$, $n=1$ in the region
of $\alpha$ 
close to 1. The scaling behavior is apparent and the numerical fit resulted in
a value $0.51 \pm 0.01$ for the slope, in accordance with expression
(\ref{sca}).}
\label{f3}
\end{figure}

For model A, we find that the force between the rings becomes attractive for 
short 
distances and $\alpha>1$. For the other cases the
force is always repulsive. As $x \to 0$, the force is finite only for
$\alpha=1$. 
\begin{figure}
\centerline{\epsfig{file=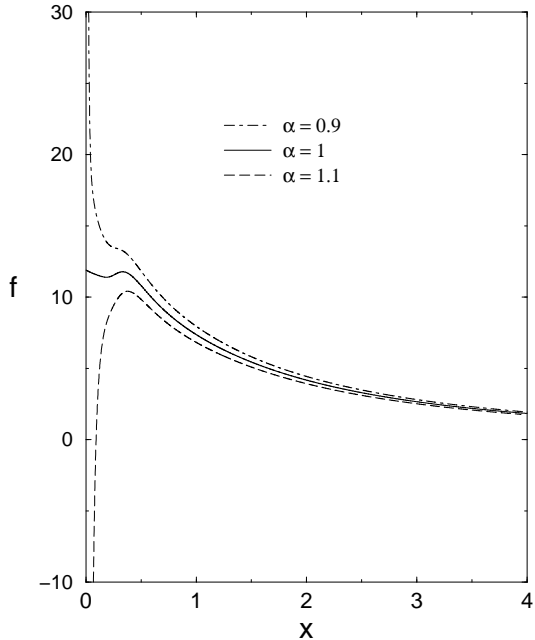,width=7cm} }
\caption{The force between the rings as a function of the distance $x$ for
model A. The
results presented are for $Z=10$, $n=2$ and $\xi=2$.}
\label{f2}
\end{figure}
The finite repulsive force observed in the limit $x \to 0$ when $\alpha=1$ may
be easily understood. In this limit, the probability of having no charges
condensed on the pair of sites $(0,1)$ and $(0,2)$ vanishes, since this would
lead to a
divergence in the energy. Thus this pair of sites does not contribute at all
to the force, and the contributions of the other pairs is either positive or
zero. This argument also explains the decreasing value of the force at vanishing
distance as $n$ is increased, as may be seen in figure \ref{f3}. 

To study behavior of the attractive force for model A at short distances
($\alpha >1$), we 
consider the limit of small $x$. In this case, we have  
\begin{equation}
\lim_{x \to 0}z_0=0,
\end{equation}
where $z_0=\exp{-\xi^\prime/x}$, is related to the interaction between the 
charges in 
positions $(0,1)$ and $(0,2)$. All the other variables $y_i$ and $z_i$ are
finite. We can therefore define the variable 
\begin{equation}
v=\min_{i}(v_A(i,0)),
\end{equation}  
where the index 0 refers to the pair of sites $(0,1)$ and $(0,2)$. We then
rewrite the partition function as 
\begin{equation}
Q_A=z_0^v\,W\,(1-P),
\end{equation}
where
\begin{equation}
W=\sum_{i=1}^l \prod_{j=1}^{Z/2}y_j^{u_A(i,j)}\prod_{k=1}^m z_k^{v_A(i,k)},
\end{equation}
and
\begin{equation}
P=\frac{1}{W}\sum_{i=l+1}^{N_c} z_0^{v_A(i,0)-v}\prod_{j=1}^{Z/2}y_j^{u_A(i,j)}
\prod_{k=1}^m z_k^{v_A(i,k)}.
\end{equation}
In these equations, we considered the first $l$ configurations to have an
exponent 
$v_A(i,0)=v$. $P$ vanishes when $x \to 0$ since $v_A(i,0)>v$ for 
$i \le l+1$, so we may approximate
\begin{equation}
Q_A \approx z_0^v\,W.
\end{equation}

The possible values of $v_A(i,0)$, 
are $1$ if there are no condensed counterions in both sites, $1-\alpha$ 
when there is one condensed counterion on one of the sites, and $(1-\alpha)^2$
if there is a condensed counterion on each site. As we discussed before,
the force becomes attractive only for $\alpha>1$, so we can conclude that the
minimum value is 
$v=1-\alpha$. Using the above approximation for the partition function 
we obtain
\begin{equation}
f=\frac{v}{x^2}+\frac{1}{\xi^\prime W}\frac{\partial W}{\partial x}.
\end{equation}
Taylor expanding the function $g(x)=\frac{1}{\xi^\prime W}\frac{\partial 
W}{\partial
x}$, $g(x) \sim g_0+g_1x+\ldots$ and keeping only the first term, we get
\begin{equation}
f=\frac{v}{x^2}+g_0.
\end{equation}
For the distance $x_0$ of vanishing force, we may then find the following
scaling relation
\begin{equation}
x_0 \approx \left(\frac{\alpha -1}{g_0}\right)^{1/2}.
\label{sca}
\end{equation}

In figure \ref{f3} curves of $x_0$ as a function of $\alpha$ are shown for
some values of $Z$, $n$, and $\xi$, and the collapse of these curves at small
values of $x_0$ may be seen. Also, in the inset the limiting behavior found in
equation (\ref{sca}) is evident.

We proceed showing results for model B. In this case, it should be stressed
that only in the rather unphysical limit of infinite temperature we have
divergence in the force at vanishing values of $x$. These divergences are
present for finite temperatures both in model A and in the similar model for
rodlike polyions 
\cite{are99,sla02} and are due to pairs of sites with nonzero charge with
vanishing distance between them. The absence of these divergences in model B
prevents us from performing 
an asymptotic analysis similar to the one done for model A, which
would also prove the existence (or not) of attractive forces in model
B. Simulations performed in a cylindrical version of model B found attractive
forces at sufficiently low separation and enough condensed counterion
charges \cite{dl01}. 

\begin{figure}
\centerline{\epsfig{file=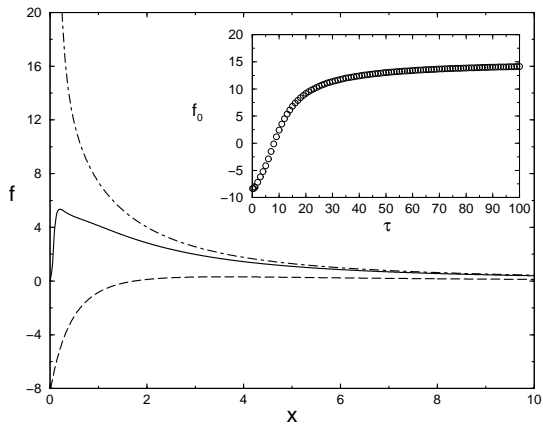,width=7cm} }
\caption{The force as a function of the distance for model B. The results
shown are for $Z=10$, $n=2$, and $\alpha=1.6$. Curves for three different
values of the inverse temperature $\xi^\prime$ are shown: Dashed line -
$\xi^\prime=2$; 
full line - $\xi^\prime=1.23$; dot-dashed line - $\xi^\prime=0$. In the inset, values
of the force at vanishing $x$ ($f_0=\lim_{x \to 0} f$) are shown as a function
of the reduced temperature $\tau=1/\xi^\prime$.} 
\label{f4}
\end{figure}

In figure \ref{f4} the force between the rings for model B is shown as a
function of the separation $x$. For sufficiently low temperatures and not too
low or too high values of $\alpha$, attractive forces are found for low
separation. As 
mentioned above, for finite temperatures the forces are always finite, since
in this model the energy of the configurations where the pair of sites $(0,1)$
and $(0,2)$ is occupied diverges as $x \to 0$. The force at vanishing
separation $f_0=\lim_{x \to 0} f(x)$ is shown as a function of the reduced
temperature $\tau=1/\xi^\prime$ in the inset of the figure, and increases
monotonically, so that, as 
expected, it diverges to $+\infty$ as the temperature becomes infinite
($1/\xi^\prime \to \infty$). We obtained similar 
results for other values of $Z$ and $n$, so that the data in figure \ref{f4}
are an example of a general trend.

\begin{figure}
\centerline{\epsfig{file=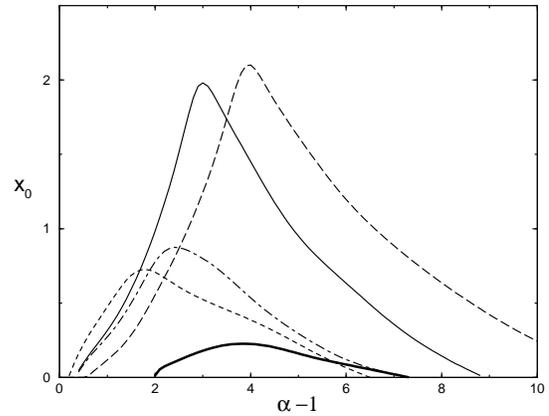,width=7cm} }
\caption{The equilibrium distance $x_0$ as a function of $\alpha-1$ for
different values of $Z$, $n$, and $\xi^\prime$ (model B). Full thin line -
$Z=8$, $n=2$, and $\xi^\prime=1$; full thick line - $Z=8$, $n=2$, and
$\xi^\prime=0.05$; long dashed line - $Z=10$, $n=2$, and $\xi^\prime=1$;
short dashed line - $Z=8$, $n=3$, and $\xi^\prime=1$; dot-dashed line - $Z=10$,
$n=3$, and $\xi^\prime=1$.}
\label{f5}
\end{figure}

Figure \ref{f5} shows results for the equilibrium distance $x_0$ as a function
of the valence $\alpha$ of the counterions for some choices of $Z$, $n$, and
$\xi^\prime$ . As
mentioned above, we were not able to perform a simple analytical study of
model B in the limit of vanishing $x$, and as expected no simple scaling
behavior is found for model B in this limit, similar to the ones observed for
model A and the other model proposed for rodlike polyions
\cite{are99,sla02}. Thus, the minimum value of $\alpha$ for the existence of
an attractive interaction varies with $Z$, $n$, and $\xi^\prime$, and so
does the initial slope of the curve $x_0 \times \alpha$. Another feature of
these curves is the existence of a {\em maximum} value of the valence of the
counterions, above which no attractive force is found. Thus, as is apparent in
figure \ref{f5}, for each curve there are two values of $\alpha$ for which
$x_0=0$. Results for $f_0$ as a function of $\alpha$ which will not be
presented here for brevity
also show this feature: The force is attractive in the interval between the
limiting values of $\alpha$, and therefore passes through a minimum in some
point inside the interval. It should be stressed, however, that the larger
limiting value of $\alpha$ we found for model B assumes values which are too
large to be physically realizable, but since the model itself is rather
artificial, one may wonder of this effect may not be seen in more realistic
models such as the one discussed in \cite{dl01} and in experiments. For model
A and the model which was proposed for rodlike polymers, no such maximum value
of $\alpha$ is found, and the curves of $x_0 \times \alpha$ pass through a
maximum but approach zero asymptotically as $\alpha \to \infty$.

\section{Final comments and discussions}
\label{III} 

In an appropriate range of the parameters, both models studied here display
attractive interaction forces, which may be explained by correlations between
the counterions condensed on the ringlike polyions. As the temperature is
increased, in general the correlations become weaker and therefore the
attractive force decreases. In simplified models which were exactly solved
before \cite{are99,sla02}, as well as in model A in this work, the attractive
forces increase without bond as the distance between the polyions is lowered,
since one or more pairs of sites with opposite charges are separated by a
distance which vanishes. This effect allowed an exact analysis of the models
at vanishing separation, but also has the drawback of being not very
reasonable physically. In particular, in the case of model A, it leads to the
conclusion that it is sufficient to have only one condensed counterion on each
polyion, with a valence slightly larger than one, to ensure the appearance of
attractive forces. As was already found in simulations of cylindrical polyions
\cite{dl01} and other more realistic models for DNA segments \cite{kor99},
attraction between 
like charged polyions is not restricted to this rather artificial situation
where sites with opposite charges become very close. In model B we find another
example of this kind: The forces are
finite for finite temperature, and a valence which is larger than one by a
finite amount is needed to ensure attractive forces. 

In both models studied here, when the parameters are
such that attractive forces are found, the equilibrium distance $x_0$ as a
function of $\alpha$ has a maximum. At low temperature, the maximum
is sharp and located close to $\alpha=Z/n$, which corresponds to
neutral polyions. As the temperature is increased, the maximum becomes broader
and 
moves towards larger values of $\alpha$. With increasing temperature, the
region of attractive force in 
the $\alpha \times x$ plane shrinks and finally disapears as the temperature
becomes infinite. An example of this behavior is seen in the two curves for
different temperatures for model B with $Z=8$ and $n=2$ in figure
\ref{f5}. Also, in figure \ref{f4} one notices that for infinite temperature
($\xi^\prime=0$), the force is always repulsive.

\acknowledgements

It is a pleasure to acknowledge Yan Levin for many suggestions in this work
and for a critical reading of the manuscript. Partial financial support by the
brazilian agencies CNPq and FAPERJ is gratefully acknowledged.

\end{multicols}

\begin{references}

\bibitem{gc2000}D. G. Grier and J. C. Crocker, Phys Rev. E {\bf61}, 
980 (2000).

\bibitem{lbt98}Y. Levin, M. C. Barbosa, and M. N. Tamashiro, 
Europhys. Lett. {\bf 41}, 123 (1998); M. N. Tamashiro, Y. Levin, and 
M. C. Barbosa, Physica A {\bf 258}, 341 (1998).

\bibitem{rh97}R. van Roij and J. P. Hansen, Phys. Rev. Lett. {\bf79}, 
3082 (1997); R. van Roij, M. Dijkstra, and P. Hansen, Phys. Rev. E 
{\bf59}, 2010 (1999).

\bibitem{dbl01}A. Diehl, M. C. Barbosa, and Y. Levin, Europhys. Lett. 
{\bf 53}, 86 (2001).

\bibitem{ll99}P. Linse and V. Lobaskin, Phys. Rev. Lett. {\bf83}, 
4208 (1999); J. Chem. Phys. {\bf112}, 3917 (2000).

\bibitem{sr90}M. J. Stevens and M. O. Robbins, Europhys. Lett. 
{\bf12}, 81 (1990); I. Rouzina and V. A. Bloomfield J. Phys. Chem. 
{\bf100}, 9977 (1996); Y. Levin, Physica A {\bf 265}, 432 (1999); A. 
Diehl, M. N. Tamashiro, M. C. Barbosa, and Y. Levin, Physica A {\bf 
274}, 433 (1999).

\bibitem{blo91} V.A. Bloomfield, Biopolymers {\bf 31}, 1471 (1991);
V.A. Bloomfield, Biopolymers {\bf 44}, 269 (1997).

\bibitem{ha97} B.-Y. Ha and A.J. Liu, {\it Phys. Rev. Lett.} 
{\bf 79}, 1289 (1997).


\bibitem{rou96} I. Rouzina and V. Bloomfield {\it J. Phys. Chem.} {\bf 100},
        9977 (1996); B.I. Shklovskii, Phys. Rev. Lett. {\bf 82}, 3268 (1999).  

\bibitem{are99} J. J. Arenzon, J. F. Stilck, and Y. Levin, Eur. Phys.
J. B {\bf 12}, 79 (1999).

\bibitem{sol99} F. J. Solis, and M. Olvera de la Cruz, Phys. Rev. E {\bf 60}, 
4496 (1999).

\bibitem{kor99} A. A. Kornyshev and S. Leikin, Phys. Rev. Lett. {\bf 82}, 4138 
(1999).

\bibitem{lev99} Y. Levin, J. J. Arenzon, and J.F. Stilck,
     Phys. Rev. Lett. {\bf 83}, 2680 (1999). 

\bibitem {man69} G. S. Manning, J. Chem. Phys. {\bf 51},  924 (1969);
Q. Rev. Biophys. {\bf 11}, 179 (1978).

\bibitem{lb97}Y. Levin and M. C. Barbosa, J. Phys. II (France) {\bf 7}, 37
(1997). 

\bibitem{dl01}A. Diehl, H. A. Carmona, and Y. Levin, Phys. Rev E {\bf 64},
1804 (2001).

\bibitem{sla02}J. F. Stilck, Y. Levin and J. J. Arenzon, J. Stat. Phys (in
press) (2002).

\end{references}
\end{document}